\documentclass[letterpaper,titlepage,11pt]{article}
\usepackage{hyperref}
\usepackage[pdftex]{graphicx}
\usepackage{epsfig}
\usepackage{graphicx, subfigure}
\usepackage{float}
\usepackage{amssymb,amsmath,amsfonts}

\def\clock{{\count0=\time
           \divide\count0 60
           \ifnum\count0<10 0\fi\the\count0
           \multiply\count0 -60 \advance\count0 \time
           :\ifnum\count0<10 0\fi \the\count0
         }}
\newcommand{\timestamp}{{\small\vbox{\hbox{\tt\jobname.tex}
\hbox{\the\day/\the\month/\the\year, \clock}}}}

\newcommand{\spa}{\ , \ \ }

\newcommand{\be}{\begin{eqnarray}}
\newcommand{\ee}{\end{eqnarray}}
\newcommand{\beq}{\begin{eqnarray}}
\newcommand{\eeq}{\end{eqnarray}}

\newcommand{\beqa}{\begin{eqnarray}}
\newcommand{\eeqa}{\end{eqnarray}}

\let\oldsqrt\sqrt
\def\sqrt{\mathpalette\DHLhksqrt}
\def\DHLhksqrt#1#2{%
\setbox0=\hbox{$#1\oldsqrt{#2\,}$}\dimen0=\ht0
\advance\dimen0-0.2\ht0
\setbox2=\hbox{\vrule height\ht0 depth -\dimen0}%
{\box0\lower0.4pt\box2}}

\numberwithin{equation}{section}

\begin{document}

\begin{titlepage}

\topmargin=-1true mm

 \vskip 3 cm

\centerline{\Huge \bf  Relativistic Elasticity}
\vskip 0.5cm
\centerline{\Huge \bf  of Stationary Fluid Branes}
\vskip 1.5cm

\centerline{\large {\bf Jay Armas} and {\bf Niels A. Obers}}

\vskip 1cm

\begin{center}
\sl The Niels Bohr Institute, University of Copenhagen\\
\sl  Blegdamsvej 17, DK-2100 Copenhagen \O , Denmark
\end{center}
\vskip 0.2cm

\centerline{\small\tt jay@nbi.dk, obers@nbi.dk}

\vskip 3cm \centerline{\bf Abstract} \vskip 0.2cm \noindent
Fluid mechanics can be formulated on dynamical surfaces of arbitrary co-dimension embedded in a background space-time. This has been the main object of study of the blackfold approach in which the emphasis has primarily been on stationary fluid configurations. Motivated by this approach we show under certain conditions that a given stationary fluid configuration living on a dynamical surface of vanishing thickness and satisfying locally the first law of thermodynamics will behave like an elastic brane when the surface is subject to small deformations. These results, which are independent of the number of space-time dimensions and of the fluid arising from a gravitational dual, reveal the (electro)elastic character of (charged) black branes when considering extrinsic perturbations.

\end{titlepage}

\small
\tableofcontents
\normalsize
\setcounter{page}{1}

\section{Introduction}
Recently, the interest in the study of fluid dynamics has increased considerably from the point of view of gravitational physics and holography. This has been triggered by different approaches to phenomena encountered in black hole physics leading to several long-wavelength effective descriptions. The membrane paradigm describes the behavior of black hole horizons when subject to certain perturbations \cite{Damour:1978cg, Parikh:1997ma, Thorne:1986, Eling:2009sj, Bredberg:2010ky}. In the context of AdS/CFT, the fluid/gravity correspondence \cite{Bhattacharyya:2008jc, Kovtun:2004de} provides an effective description for the behavior of perturbations of black branes along boundary directions. The blackfold approach \cite{Emparan:2007wm, Emparan:2009cs, Emparan:2009at, Camps:2010br, Camps:2012hw}, an effective theory for the dynamics of thin black branes, characterizes the behavior of strained black brane geometries along worldvolume directions and orthogonal to these All these approaches show that the physical properties of black holes in certain regimes can be captured by the physics of fluid flows that either live on the stretched horizon (membrane paradigm), on the boundary (fluid/gravity) or in an intermediate region (blackfold approach). These insights motivate the search for a generalization of the mechanics of fluid flows that live on an arbitrary surface of any co-dimension placed in a background space-time. 

This development has taken place within the realm of the blackfold approach \cite{Emparan:2009at, Emparan:2011hg, Caldarelli:2010xz}, and has lead to a vast study of large classes of fluids with particular focus on stationary configurations. To make the reader familiar with this effective theory we present the equations of motion that describe it, first derived by Carter \cite{Carter:2000wv},
\beq \label{bf_eqs} 
D_{a}T^{ab}=0 \quad , \quad T^{ab}{K_{ab}}^{\rho}=0~.
\eeq 
Here $D_{a}$ is the covariant derivate with respect to the induced metric $\gamma_{ab}$ of the surface on which the fluid lives while ${K_{ab}}^{\rho}$ is its extrinsic curvature. When the stress-energy tensor $T^{ab}$ is assumed to be of the perfect fluid form, the first set of equations give rise to the usual energy density continuity and Euler equations of a perfect fluid. When considering, for example, the type of (intrinsic) hydrodynamic perturbations of black hole horizons encountered in the context of the fluid/gravity correspondence, this set of equations can be derived as constraint equations directly from Einstein equations \cite{Bhattacharyya:2008jc}. One can think of perturbations of this type as fluctuations in the fields that characterize the material - in this case the fluid - that lives on a space-time surface. Proceeding order-by-order in perturbation theory results in dissipative corrections to the stress-energy tensor. The second set of equations in \eqref{bf_eqs} is of extrinsic nature and is associated with deformations of the geometry (surface) on which the fluid flows. As in the case of hydrodynamic fluctuations,  this set is also directly derivable as constraint equations from Einstein equations \cite{Emparan:2007wm, Camps:2012hw} when deformations, for example bending, of black brane geometries are considered. Our aim in this note is to show that this extrinsic set of equations can be viewed as a relativistic generalization of elasticity theory of thin branes.

 The reader may be familiar with Eqs.~\eqref{bf_eqs} when dealing with Dirac branes where $T^{ab}=T_{D_p}\gamma^{ab}$ with $T_{D_p}$ being the tension of the $p$-brane but we note that they have a direct non-relativistic analog when considering deformations of thin membranes. To be precise, suppose that we are given a thin elastic membrane of thickness $r_0$ and subject it to external forces applied at its circumference causing it to stretch in all directions. Assuming the material to behave elastically, generating internal stresses $\sigma^{ab}$ that encode its Hookean response to the stretching, the equations of motion that govern its mechanical equilibrium can be obtained by varying the free energy \cite{Landau:1959te}
 \beq \label{F_class}
 F[X^{\mu}]=\frac{1}{2}\int_{Vol}dV\sigma^{ab}U_{ab}~,
 \eeq   
 where $X^{\mu}$ is the set of mapping functions that parametrize the position of the membrane in the ambient space while $U^{ab}$ is the strain tensor. The resulting set of equations is exactly \eqref{bf_eqs} with $T^{ab}$ replaced by $\sigma^{ab}$ and with the indices $a,b$ only running through the spatial directions. Within this perspective, a response of the Dirac brane type can be seen as the isotropic stretching of a $p$-brane.
 
The motivation for this interpretation has its origin in the analysis of the corrections associated with extrinsic perturbations of black branes performed in \cite{Armas:2011uf, Camps:2012hw, Armas:2012ac} which takes into account the finite thickness of the brane geometry. In fact, for the particular case of black strings bent into circular shape, Eqs.~\eqref{bf_eqs} keep the same form, with the stress-energy tensor acquiring a correction that can be interpreted as a Hookean elastic response of the string due to the bending \cite{Armas:2011uf, Armas:2012ac}. In line with this, we ought to show that the extrinsic dynamics is already to leading order that of an elastic solid. Even though greatly inspired by black brane physics the scope of the results that will be presented here is far wider. As a matter of a fact, what will be shown is this: given a stationary fluid configuration satisfying local thermodynamic laws living on a thin space-time surface, assuming diffeomorphism invariance and ignoring backreaction, its extrinsic dynamics in transverse directions to the surface correspond to that of an elastic brane. We expect this interpretation to lead in the future to a formal development of general relativistic elasticity theory of branes that can be captured from gravity in the same spirit as a rigorous development of fluid and superfluid dynamics was captured using the fluid/gravity correspondence (see \cite{Banerjee:2008th, Erdmenger:2008rm, Son:2009tf, Banerjee:2008th, Herzog:2011ec, Bhattacharya:2011tra}). Refs. \cite{Armas:2011uf, Camps:2012hw, Armas:2012ac} already made important steps towards this goal by measuring for the first time response coefficients, such as elastic and piezoelectric moduli, of materials - black branes - bent both in space as well as in time.

This note is organized as follows. In Sec.~\ref{fluid_dyna} we review and reinterpret the equations of motion for stationary fluids living on dynamical surfaces \cite{Emparan:2009cs, Emparan:2009at, Emparan:2011hg, Caldarelli:2010xz} in terms of elastic solid concepts. We show this to be the case for isotropic fluids that in addition may carry different types of charges. This solid interpretation along extrinsic (transverse) directions to the embedding surface also holds for the anisotropic fluids of \cite{Emparan:2011hg, Caldarelli:2010xz} carrying string charge. In Sec.~\ref{bf_branes} we give examples of fluid configurations arising from gravitational duals satisfying these properties and apply the formalism with this new interpretation obtaining the elastic equilibrium condition for certain black holes. In Sec.~\ref{discussion} we conclude.

\section{Stationary fluids on dynamical surfaces}  \label{fluid_dyna}
In this section we review and reinterpret the theory describing stationary fluid configurations confined to a $(p+1)$-dimensional dynamical surface, parametrized by the mapping functions $X^{\mu}$ embedded in an ambient $D$-dimensional space-time with metric $g_{\mu\nu}$ \cite{Emparan:2007wm, Emparan:2009cs}. When the surface and the space-time are of equal dimension we obtain the ordinary description of fluid mechanics. We work under the assumption that the fluid is in local thermodynamic equilibrium. This is achieved when the mean free path characterized by the inverse of the local temperature $\mathcal{T}(\sigma^a)$ is much smaller than the radius of curvature of the embedding geometry $R(\sigma^{a})$, i.e., 
\beq \label{thermo_eq}
\frac{1}{\mathcal{T}(\sigma^{a})}\ll R(\sigma^{a})~,
\eeq
where $\sigma^{a},~a=0,...,p$ are the coordinates that parametrize the worldvolume $\mathcal{W}_{p+1}$ traced out by the surface in space-time and endowed with metric $\gamma_{ab}=g_{\mu\nu}\partial_a X^{\mu}\partial_b X^{\nu},~\mu,\nu=0,...,D-1$, with Lorentzian signature. We further assume that the surface is infinitely thin. Finite thickness effects have been considered in \cite{Armas:2011uf, Armas:2012ac}. However we note that as in the classical example discussed around Eq.~\eqref{F_class} where the thickness can be integrated out and the object replaced by an infinitely thin elastic membrane, the thickness of the $(p+1)$-dimensional objects considered here can also be integrated out giving rise to a worldvolume effective theory. Secs.~\ref{wveff} and \ref{intfluid} review this theory and the intrinsic fluid dynamics following Refs.~\cite{Emparan:2009cs, Emparan:2009at, Caldarelli:2010xz, Emparan:2011hg}, while the remaining reinterpret the extrinsic dynamics in terms of relativistic elasticity theory.

\subsection{The worldvolume effective theory} \label{wveff}
Assuming the fluid not to backreact onto the background, the usual equations of fluid mechanics can be derived by imposing conservation of the stress-energy tensor $\hat{T}^{\mu\nu}$ \cite{Carter:2000wv}
\beq \label{st_conserv}
\nabla_{\nu}\hat{T}^{\mu\nu}=\hat{\mathcal{F}}^{\mu}~,
\eeq
where we have included the possibility of an external force $\hat{\mathcal{F}}^{\mu}$. The dynamics of fluids living on dynamical surfaces also follows from \eqref{st_conserv} but the stress-energy tensor characterizing these configurations is confined to the surface in the following way:
\beq \label{st_thin}
\hat{T}^{\mu\nu}(x^{\alpha})=\int_{\mathcal{W}_{p+1}}d^{p+1}\sigma\sqrt{-\gamma}T^{\mu\nu}(\sigma^{a})\frac{\delta^{D}(x^{\mu}-X^{\mu}(\sigma^{a}))}{\sqrt{-g(x^{\alpha})}}~.
\eeq
Here one should interpret $\hat{T}^{\mu\nu}(x^{\alpha})$ as the full spacetime stress-tensor while the components $T^{\mu\nu}(\sigma^{a})$ with support on $\mathcal{W}_{p+1}$ should be seen as worldvolume densities of stress-energy. We assume the force term $\hat{\mathcal{F}}^{\mu}(x^{\alpha})$ to be of the same form as \eqref{st_thin} with appropriate worldvolume density $\mathcal{F}^{\mu}(\sigma^{a})$. Introducing \eqref{st_thin} into Eq.~\eqref{st_conserv} results in a worldvolume theory where the densities $T^{\mu\nu}(\sigma^{a})$ only have components tangential to the worldvolume
\beq \label{cons_thin}
{\perp^{\rho}}_{\nu}T^{\mu\nu}=0~,
\eeq
where the orthogonal projector to the worldvolume is defined as $\perp^{\rho\nu}=g^{\rho\nu}-\gamma^{\rho\nu}$ with $\gamma^{\mu\nu}=\gamma^{ab}\partial_{a}X^{\mu}\partial_b X^{\nu}$ being the push forward of the induced metric. The equations of motion derivable from Eq.~\eqref{st_conserv} can be written in the form
\beq \label{dyn_thin}
{\gamma^{\rho}}_{\nu}\nabla_{\rho}T^{\mu\nu}=\mathcal{F}^{\mu}~,
\eeq
together with the boundary condition
\beq \label{bc_thin}
T^{\mu\nu}\hat{n}_{\nu}|_{\partial \mathcal{W}_{p+1}}=0~,
\eeq
where $\hat{n}_{\nu}$ is a unit normal vector orthogonal to the worldvolume boundary. Imposing the constraint \eqref{cons_thin} on $T^{\mu\nu}$ requires $T^{\mu\nu}=T^{ab}u_{a}^{\mu}u_{b}^{\nu}$ with $u_a^{\mu}=\partial_a X^{\mu}$ which can then be used to split Eq.~\eqref{dyn_thin} into two sets of equations by projecting along tangential and orthogonal directions to the worldvolume:
\beq \label{bfe_int}
D_{b}T^{ab}=u^{a}_{\mu}\mathcal{F}^{\mu}~,
\eeq
\beq\label{bfe_ext}
T^{ab}{K_{ab}}^{\rho}=\perp^{\rho}_{\mu}\mathcal{F}^{\mu}~.
\eeq
These equations reduce to Eqs.~\eqref{bf_eqs} when $\mathcal{F}^{\mu}=0$. The first equation expresses the conservation of worldvolume stress-energy while the second can be interpreted as the balance of forces acting on the fluid in orthogonal directions.  We will in the remaining sections take $T^{ab}$ to be of the perfect fluid form and $\mathcal{F}^{\mu}$ to be vanishing but we note that this worldvolume effective theory is valid for any type of material one would like to describe.

\subsection{Intrinsic fluid dynamics} \label{intfluid}
Our case study is that of a perfect fluid with energy density $\epsilon(\sigma^{a})$ and pressure $P(\sigma^{a})$ with associated local entropy $s(\sigma^{a})$ and corresponding local temperature $\mathcal{T}(\sigma^{a})$. The stress-energy tensor is given by
\beq\label{0stress}
T^{ab}=(\epsilon+P)u^{a}u^{b}+P\gamma^{ab}~.
\eeq
Here the fields $u^{a}$ denote the fluid velocities normalized such that $u^{a}u_{a}=-1$. Assuming local thermodynamic equilibrium the first law of thermodynamics must be satisfied
\beq \label{de}
d\epsilon=\mathcal{T}ds~,
\eeq
where the infinitesimal differentials are taken along worldvolume directions. In addition these quantities are supplemented by the Gibbs-Duhem relations
\beq \label{dp}
\epsilon+P=\mathcal{T}s~,~dP=sd\mathcal{T}~.
\eeq
The fluid dynamical equations follow directly from the intrinsic equation \eqref{bfe_int}, which upon contraction along directions tangential and orthogonal to the fluid flows and using the above local thermodynamic relations can be written as the conservation of the entropy current $J^{a}_{s}$,
\beq \label{entropy_conser}
D_{a}J_{s}^{a}=0~,~J^{a}_{s}=su^{a}~,
\eeq
and the Euler force equations
\beq \label{Euler_force}
\hat{\perp}^{ab}\mathcal{T}s(\dot{u}_b+\partial_b\text{ln}\mathcal{T})=0~,
\eeq
where $\hat{\perp}^{ab}=\gamma^{ab}+u^{a}u^{b}$. Eqs.~\eqref{entropy_conser}-\eqref{Euler_force} are subject to the boundary condition \eqref{bc_thin} now in the form:
\beq
T^{ab}u_{a}^{\mu}\hat{n}_{b}|_{\partial \mathcal{W}_{p+1}}=0~.
\eeq
Focusing on stationary fluid configurations, these two equations are solved by requiring the fluid velocities $u^{a}$ to be proportional to a worldvolume Killing vector field $\mathbf{k}$ of the form \cite{Caldarelli:2008mv}
\beq \label{killing_vector}
\mathbf{k}=\xi+\Omega_{i}\chi_{i}~,
\eeq
where $\xi$ is a worldvolume timelike Killing vector, $\chi_{i}$ are spacelike ones and $i$ labels spatial worldvolume directions. Without loss of generality we choose
\beq \label{u}
u^{a}=\frac{\mathbf{k}^{a}}{|\mathbf{k}|}~.
\eeq
The global fluid temperature $T$ appears as an integration constant from Eq.~\eqref{Euler_force} and it is related to the local temperature $\mathcal{T}$ via a local redshift factor,
\beq \label{ass_t}
T=|\textbf{k}|\mathcal{T}~,
\eeq
while the total entropy, assuming $\textbf{k}_a$ to be hypersurface orthogonal with respect to the worldvolume metric, can be obtained by integrating the entropy current over the spatial part of the worldvolume $\mathcal{B}_{p}$,
\beq \label{total_entropy}
S=-\int_{\mathcal{B}_{p}}dV_{(p)}su^{a}n_{a}~.
\eeq
Here we introduced the spatial measure $dV_{(p)}$ on the worldvolume and defined an orthogonal vector $n_{a}$ to a worldvolume spacelike hypersurface  in the manner
\beq
n^{a}=\frac{\xi^{a}}{R_0}~,
\eeq
where $R_0$ is the norm of the timelike Killing vector field $\xi$ on the worldvolume.


\subsection{Extrinsic elastic dynamics}
In this section we analyze the extrinsic dynamics of fluid configurations described by Eq.~\eqref{bfe_ext}. We begin by defining the state of strain of the brane and the strain tensor and then proceed to describe the equations of mechanical equilibrium.


\subsubsection*{The state of strain and the strain tensor}
The fluid configuration studied in the previous section lives on an infinitely thin surface described by the induced metric $\gamma_{\mu\nu}$. We are interested in examinaning the thermodynamic properties of such a fluid when the geometry is deformed along orthogonal directions. The metric $\gamma_{\mu\nu}$ measures distances between neighboring points on the embedding surface, therefore, working under the assumption that the surface is thin and hence that variations in distances measured with $\perp_{\mu\nu}$ can be ignored\footnote{This is the usual assumption of classical elasticity theory when considering deformations of thin membranes and small strains and stresses \cite{Landau:1959te}.}, the metric $\gamma_{\mu\nu}$ describes the local state of strain of the brane.

Let us define the state of strain of the brane prior to a deformation (unstrained state) by $\bar\gamma_{\mu\nu}$. The length of infinitesimal spacetime distances along the surface is thus of the form
\beq
\bar{ds}^2=\bar\gamma_{\mu\nu}dx^{\mu}dx^{\nu}~.
\eeq
After a deformation, the state of strain is no longer described by $\bar\gamma_{\mu\nu}$ but instead by the actual value of $\gamma_{\mu\nu}$, hence the length of the infinitesimal element is changed to
\beq
ds^2=\gamma_{\mu\nu}dx^{\mu}dx^{\nu}~.
\eeq
Assuming the strains and stresses involved to be small, the difference between the length of the line elements of the strained and unstrained case along an arbitrary orthogonal vector $N_{\rho}$ is given by:
\beq
\Delta s^2=ds^2-\bar{ds}^2=\pounds_{N}\gamma_{\mu\nu} dx^{\mu}dx^{\nu}~.
\eeq
Since we are only interested in changes along worldvolume directions we project this measure along those directions,
\beq
\Delta s^2|_{\mathcal{W}_{p+1}}=\gamma^{\lambda}_{\mu}\gamma^{\rho}_{\nu}\pounds_{N}\gamma_{\lambda\rho} dx^{\mu}dx^{\nu}=-2N_{\rho}{K_{\mu\nu}}^{\rho}dx^{\mu}dx^{\nu}~,
\eeq
where we have used a mathematical identity described in \cite{Emparan:2009at}. Therefore, along any orthogonal direction $N_{\rho}$, the strain varies proportionally to the extrinsic curvature tensor ${K_{\mu\nu}}^{\rho}$ \footnote{Preliminary arguments in this direction were first given in \cite{Armas:2011uf}.}. Since ${K_{\mu\nu}}^{\rho}$ satisfies the property ${K_{\mu\nu}}^{\rho}=u^{a}_{\mu}u^{b}_{\nu}{K_{ab}}^{\rho}$, we define the Lagrangian strain tensor \cite{Quintana:1972} for the brane as 
\beq \label{strain}
U_{ab}=-\frac{1}{2}\left(\gamma_{ab}-\bar\gamma_{ab}\right)~,
\eeq
which for infinitesimal deformations reads
\beq
dU_{ab}=-\frac{1}{2}d\gamma_{ab}=N_{\rho}{K_{ab}}^{\rho}~.
\eeq


\subsubsection*{Equations of extrinsic dynamics} 
The extrinsic dynamics of the fluid living on the dynamical surface are described by the extrinsic equation \eqref{bfe_ext} which can be written in the form \cite{Emparan:2009at}
\beq \label{ext}
\mathcal{T}s{\perp^{\rho}}_{\mu}\dot{u}^{\mu}+PK^{\rho}=0\spa
\eeq
relating the acceleration of the fluid along orthogonal directions to the action of the mean extrinsic curvature $K^{\rho}\equiv\gamma^{ab}{K_{ab}}^{\rho}$. Here, $\dot{u}^\mu$ is the fluid acceleration defined as $\dot{u}^{\mu}\equiv u^{\nu}\nabla_{\nu}u^{\mu}$. We now make the assumption of the existence of a background Killing vector $\textbf{k}^{\mu}$ whose pullback onto the worldvolume coincides with the worldvolume Killing vector field $\textbf{k}^{a}$. Under this assumption we can write
\beq \label{udot}
\mathcal{T}s\dot{u}_{\mu}=-s\partial_{\mu}\mathcal{T}~.
\eeq
Introducing this into Eq.~\eqref{ext} and using the definition of $K^{\rho}$ yields,
\beq
{\perp^{\rho}}_{\mu}s\partial^{\mu}\mathcal{T}=P\gamma^{ab}{K_{ab}}^{\rho}~.
\eeq
Contracting the last equation with an arbitrary orthogonal vector $N_{\mu}$ leads to
\beq \label{dext}
sd\mathcal{T}=-\frac{1}{2}\mathcal{\sigma}^{ab}d\gamma_{ab}~,
\eeq
where the infinitesimal differentials denote a variation along orthogonal directions and where we have defined the pressure tensor $\sigma^{ab}$ as:
\beq \label{pressure_tensor}
\sigma^{ab}=P\gamma^{ab}~.
\eeq
The r.h.s. of Eq.~\eqref{dext} can be written as $\sigma^{ab}dU_{ab}$ by recognizing the strain tensor defined in \eqref{strain}. We now define the elastic solid density $\rho$ as 
\beq \label{s_dens}
\rho=\epsilon+P=\mathcal{T}s\spa
\eeq
and hence interpret $s$ as an average particle density satisfying the conservation law \eqref{entropy_conser} and $\mathcal{T}$ as a mass function. In terms of infinitesimal variations along orthogonal directions to the worlvolume we find
\beq \label{dr0}
\begin{split}
d\rho&=\mathcal{T}ds + sd\mathcal{T}\\
&=\mathcal{T}ds+\sigma^{ab}dU_{ab}=\mathcal{T}ds-Pd\mathcal{V}~,
\end{split}
\eeq 
where we have used the mathematical identity obtained in \cite{Emparan:2009at} for the relative change in the local volume element
\beq
d\mathcal{V}\equiv\frac{\delta_{N}\sqrt{-\gamma}}{\sqrt{-\gamma}}=\frac{1}{2}\gamma^{ab}d\gamma_{ab}~.
\eeq
Eq.~\eqref{dr0} is exactly the relation that an elastic solid density $\rho$ should respect under hydrostatic compression \cite{Landau:1959te} and expresses the fact that the system described by this set of equations accounts for only changes in volume but not in shape. This is due to the assumption of an infinitely thin surface since effects due to bending or torsion would require a varying concentration of material across transverse directions. Along orthogonal directions we can thus write the equivalent relations
\beq \label{eq_rho}
\left(\frac{\partial \rho}{\partial U_{ab}}\right)_{s}=\sigma^{ab},~\left(\frac{\partial \rho}{\partial \mathcal{V}}\right)_{s}=-P~.
\eeq
Here, as well as in \eqref{dr0}, we have assumed that all strain components $dU_{ab}$ are linearly independent. This is not necessarily the case and we will deal with linear dependence towards the end of this section. We note that the first equality in \eqref{eq_rho} only constrains the components of the pressure tensor which are contracted with non-vanishing components of the strain tensor in \eqref{dext}. In order to make further contact with relativistic elasticity theory \cite{Quintana:1972, Karlovini:2002fc, Beig:2002pk} note that by means of the definition \eqref{s_dens} the above thermodynamic quantities and stresses can be derived from the mass function $\mathcal{T}$, for example
\beq
s\left(\frac{\partial \mathcal{T}}{\partial \mathcal{V}}\right)=-P~.
\eeq
Furthermore, we rewrite the stress-energy tensor \eqref{0stress} as
\beq \label{pstress}
T^{ab}=\rho u^{a}u^{b}+\sigma^{ab}~,
\eeq
motivating the interpretation of a solid at rest which has suffered hydrostatic compression along all worldvolume directions. The definition and interpretation of Eq.~\eqref{s_dens} together with the relations \eqref{dr0} and the decomposition \eqref{pstress} is one of the central results in this note as they express the elastic character of the relativistic material since the components of the pressure tensor $\sigma^{ab}$ involved in the extrinsic dynamics \eqref{dext} can be obtained from a single potential $\rho$ - a required condition in relativistic elasticity theory \cite{PhysRevD.1.1013, Quintana:1972, Sokolov:1994mv, Beig:2002pk, Sokolov:2003}. It is possible to consider a more general elastic potential which is also better suited when we consider charged branes. This will be the aim of the next section. For now, we focus on the case for which the strain components in \eqref{dr0} can be linearly dependent. In such situations, given the independent components of the strain tensor $U_{\tilde{a} \tilde{b}}$ we write \eqref{dr0} as
\beq\label{dr1}
d\rho=\mathcal{T}ds+\tilde\sigma^{\tilde a \tilde b}dU_{\tilde a \tilde b}~,
\eeq
where we have introduced the effective pressure tensor along the component $(\tilde a,\tilde b)$,
\beq\label{sigmatilde}
\tilde\sigma^{\tilde a \tilde b}={\perp^{\tilde a \tilde b}}_{ab}\sigma^{ab}~,\quad~{\perp^{\tilde a \tilde b}}_{ab}=\frac{\partial \gamma_{ab}}{\partial \gamma_{\tilde a \tilde b}}~.
\eeq
Here the operator ${\perp^{\tilde a \tilde b}}_{ab}$ acts as a projector onto the linearly independent subspace of the components of strain. Using Eqs.~\eqref{dr1}-\eqref{sigmatilde} we can write
\beq \label{effrho}
\left(\frac{\partial\rho}{\partial U_{\tilde a \tilde b}}\right)_s =\tilde \sigma^{\tilde a \tilde b}~.
\eeq
We wish to rewrite an effective form for the stress-energy tensor \eqref{pstress} along the directions $(\tilde a,\tilde b)$. To this aim we note that from Eqs.~\eqref{ext}-\eqref{udot} together with the identity $u^{\mu}u^{\nu}{K_{\mu\nu}}^{\rho}={\perp^{\rho}}_{\mu}\dot{u}$ \cite{Emparan:2009at} and the Gibbs-Duhem relation \eqref{dp} one finds the expression
\beq \label{forces}
\rho u^{a}u^{b}=2\left(\frac{\partial P}{\partial \gamma_{ab}}\right)~,
\eeq
which is only valid along transverse directions. Acting with the projector ${\perp^{\tilde a \tilde b}}_{ab}$ on both sides of \eqref{forces} and using \eqref{sigmatilde} we can write the effective stress-energy tensor along transverse directions as
\beq \label{effectivestress}
\tilde T^{\tilde a \tilde b}=2\left(\frac{\partial P}{\partial \gamma_{\tilde a \tilde b}}\right)+\tilde \sigma^{\tilde a \tilde b}~,
\eeq
which must satisfy the constraint:
\beq \label{const}
\tilde T^{\tilde a \tilde b}=0~.
\eeq
Imposing \eqref{const} leads directly to the extrinsic equations of motion \eqref{ext}.


\subsection{Elastic free energy, elasticity tensor and conserved charges} \label{act_princ_1}
Using the identity \eqref{dp} it is possible to rewrite Eq.~\eqref{dext} only in terms of the pressure $P$ as
\beq \label{dp0}
dP=-\frac{1}{2}\mathcal{\sigma}^{ab}d\gamma_{ab}~,
\eeq
or alternatively, along orthogonal directions and for the independent components of $d\gamma_{\tilde a \tilde b}$,
\beq \label{ext0}
-2\left(\frac{\partial P}{\partial \gamma_{\tilde a \tilde b}}\right)=\tilde\sigma^{\tilde a \tilde b}~.
\eeq
One can view this equation as a balance of forces between the pressure tensor and the internal stresses generated by a variation in volume\footnote{For the fluid branes arising from a gravitational dual analyzed in \cite{Emparan:2009at} the pressure tensor takes the interpretation of gravitational tension acting as a compression force while the first term in Eq.~\eqref{ext0} takes the interpretation of a centripetal force acting outwards the surface when the fluid is rotating. This is clear from the r.h.s. of \eqref{forces} since it is proportional to a density $\rho$ and two copies of the four-velocity $u^{a}$.}. Moreover, at the level of uncharged fluid branes, Eq.~\eqref{ext0} is equivalent to Eq.~\eqref{effrho} but this is not so in general. In fact, the pressure $P$, for reasons that will become apparent, provides a more general elastic potential which from now on we take to be the canonical one. From Eq.~\eqref{dp0} it is possible to define the bulk modulus or modulus of hydrostatic compression $K$ that measures the material response to variations in volume through the relation
\beq \label{mod_rig}
\frac{1}{K}=\left(\frac{\partial \mathcal{V}}{\partial P}\right)_{T}=-\frac{1}{P}~.
\eeq
The definition \eqref{mod_rig} has a direct classical analog \cite{Landau:1959te}. Eq.~\eqref{dp0} can be integrated to an action \cite{Emparan:2009at}
\beq \label{f0}
I[X^{\mu}]=\int_{\mathcal{W}_{p+1}}\sqrt{-\gamma}P~,
\eeq
which resembles the usual action for fluids living on a fixed background, the difference being that $\gamma$ is the volume measure on the worldvolume instead of being on the space-time \cite{Brown:1992kc}. On the other hand, the worldvolume being a surface of co-dimension higher than zero, allows for elastic behavior which can be seen from the variation of the integrand along orthogonal directions
\beq \label{dF}
\begin{split}
-d(\sqrt{-\gamma}P)&=-\sqrt{-\gamma}dP-d\left(\sqrt{-\gamma}\right)P \\
&=\sqrt{-\gamma}\left(-sd\mathcal{T}-Pd\mathcal{V}\right)~,
\end{split}
\eeq
where we have used the Gibbs-Duhem relation \eqref{dp}. Eq.~\eqref{dF} has direct analogy with the variational form of the Helmholtz free energy of an elastic solid, exhibiting the same local thermodynamic properties \cite{Landau:1959te}. The action \eqref{f0} demands to be interpreted as the solid free energy $F[X^{\mu}]\equiv-I[X^{\mu}]$ when the material has suffered hydrostatic compression. To motivate this interpretation even further we assume in the present moment the fluid to be barotropic, characterized by an equation of state $\epsilon=w P$. This together with \eqref{dp} and \eqref{s_dens} allows us to rewrite the solid free energy as
\beq \label{solid_action}
F[X^{\mu}]=-\frac{1}{w+1}\int_{\mathcal{W}_{p+1}}\sqrt{-\gamma}\rho
\eeq
which is the usual action for relativistic elastic media \cite{Salt:1971, Karlovini:2002fc, Beig:2002pk}. The assumption of barotropy can be relaxed in order to obtain an expression of the form of \eqref{solid_action} as we show in Sec.~\ref{c_fluids} when considering charged fluids. We note that the action \eqref{f0} is very general extending also to charged fluid branes \cite{Grignani:2010xm, Caldarelli:2010xz, Emparan:2011hg, Grignani:2012iw, Armas:2012bk}, while the action \eqref{solid_action} is dependent on the equation of state. For this reason we have chosen the elastic potential $P$ to be the canonical one instead of $\rho$.

\subsubsection*{The elasticity tensor}
In elasticity theory, given the potential $P$ from which the pressure tensor $\tilde{\sigma}^{\tilde a \tilde b}$ can be obtained, it is possible to define a deformation tensor $\tilde K^{\tilde a \tilde b \tilde c \tilde d}$ as the variation of the pressure tensor with respect to the state of strain \cite{Quintana:1972, Karlovini:2002fc, Beig:2002pk}. For the fluid branes considered here this has the general form
\beq \label{deformationtensor}
\tilde K^{\tilde a \tilde b \tilde c \tilde d}\equiv\left(\frac{\partial\tilde\sigma^{\tilde a\tilde b}}{\partial U_{\tilde c\tilde d}}\right)=\left(\frac{\partial^2 P}{\partial U_{\tilde a\tilde b}\partial U_{\tilde c\tilde d}}\right)~,
\eeq
satisfying the properties $\tilde K^{\tilde a \tilde b \tilde c \tilde d}=\tilde K^{(\tilde a \tilde b) (\tilde c \tilde d)}=\tilde K^{\tilde c \tilde d \tilde a \tilde b}$. Note that in the second equality in \eqref{deformationtensor} we have imposed the constraint \eqref{ext0}. Given a certain stationary fluid brane in a certain prestrained state satisfying \eqref{ext0}, one can apply a deformation taking the configuration to another strained state. The definition \eqref{deformationtensor} implies that variations of the pressure tensor along orthogonal directions are related to the deformation tensor as
\beq \label{ddeformation}
d\tilde\sigma^{\tilde a \tilde b}=\tilde K^{\tilde a \tilde b \tilde c \tilde d}dU_{\tilde c \tilde d}~.
\eeq
In order to give a concise explicit expression for \eqref{deformationtensor} we introduce an effective pressure $\tilde P(\gamma^{\tilde a \tilde b})$ along a direction $(\tilde a, \tilde b)$ such that
\beq
\tilde\sigma^{\tilde a \tilde b}=\tilde P(\gamma^{\tilde a \tilde b})\gamma^{\tilde a \tilde b}~.
\eeq
Using then the definition \eqref{deformationtensor} we obtain the general expression
\beq \label{deformation}
\tilde K^{\tilde a \tilde b \tilde c \tilde d}=-2\left(\left(\frac{\partial \tilde P}{\partial \gamma_{\tilde a\tilde b}}\right)\gamma^{\tilde c \tilde d}-\tilde P\gamma^{\tilde a(\tilde c}\gamma^{\tilde d)\tilde b}\right)~.
\eeq
This has the expected form of the deformation tensor of a material that is responding to stretching or compression signaling the fact that since the surface is taken to be infinitely thin, only variations in volume can be accounted for. The second term in \eqref{deformation} is the usual term when the pressure $\tilde P$ is constant while the first term arises due to pressure variations. For Dirac branes the first term on the r.h.s. of \eqref{deformation} vanishes expressing isotropic compression at constant pressure. 

The relativistic elasticity tensor $\tilde E^{\tilde a \tilde b \tilde c \tilde d}$ can be expressed in an analogous way with respect to the stress-energy tensor \eqref{effectivestress} \cite{Quintana:1972, Karlovini:2002fc, Beig:2002pk}. In orthogonal directions to the worldvolume, the elasticity tensor describes variations of $\tilde T^{\tilde a \tilde b}$ such that
\beq \label{dstress}
d\tilde T^{\tilde a \tilde b}=\tilde E^{\tilde a \tilde b \tilde c \tilde d}dU_{\tilde c \tilde d}~.
\eeq
According to \eqref{dstress} the elasticity tensor takes the following generic form
\beq \label{elasticity}
\tilde E^{\tilde a \tilde b \tilde c \tilde d}\equiv\left(\frac{\partial\tilde T^{\tilde a\tilde b}}{\partial U_{\tilde c\tilde d}}\right)=\left(\tilde K^{\tilde a \tilde b \tilde c \tilde d}-\left(\frac{\partial^2 P}{\partial U_{\tilde a \tilde b} U_{\tilde c \tilde d}}\right)\right)~,
\eeq
where in the second equality we have used \eqref{deformation}. The tensor  $\tilde E^{\tilde a \tilde b \tilde c \tilde d}$ satisfies the usual properties of an elasticity tensor $\tilde E^{\tilde a \tilde b \tilde c \tilde d}=\tilde E^{(\tilde a \tilde b) (\tilde c \tilde d)}=\tilde E^{\tilde c \tilde d \tilde a \tilde b}$. We note that the Eqs.~\eqref{ddeformation}, \eqref{dstress} express linear Hokean deformations of the pressure and stress-energy tensors. We will evaluate \eqref{elasticity} explicitly for neutral black branes in Sec.~\ref{bf_branes}.

\subsubsection*{Conserved charges} 
The action \eqref{f0} has a thermodynamic interpretation \cite{Emparan:2009at}. To make it precise, we note that from Eqs.~\eqref{dyn_thin} with vanishing external force one can construct a set of conserved surface currents $T^{\mu\nu}\textbf{k}_{\mu}$, such that
\beq \label{ccur}
{\gamma^{\rho}}_{\nu}\nabla_{\rho}(T^{\mu\nu}\textbf{k}_{\mu})=0~,
\eeq
where $\textbf{k}_{\mu}$ should be interpreted here as a generic space-time Killing vector field. The total energy $M$ and angular momentum $J^{i}$ of the fluid brane can then be computed by integrating the surface currents over the spatial part of the worldvolume in the following way:
\beq
M= \int_{\mathcal{B}_{p}}dV_{(p)}T^{ab}\xi_{a}n_{b}~,~J^{i}=- \int_{\mathcal{B}_{p}}dV_{(p)}T^{ab}\chi_{a}^{i}n_{b}~.
\eeq
Here we have assumed $\textbf{k}_{\mu}$ to be hypersurface orthogonal with respect to the space-time metric and also $\textbf{k}_{a}$ to be hypersurface orthogonal with respect to the worldvolume metric\footnote{Other cases where $\textbf{k}_{\mu}$ or $\textbf{k}_{a}$ are not hypersurface orthogonal have been considered in \cite{Camps:2008hb, Armas:2012bk}.}.
To proceed further, we introduce the fluid Gibbs free energy density $\mathcal{G}$ as
\beq \label{gibbs_neutral}
\mathcal{G}=\epsilon-\mathcal{T}s=-P~,
\eeq
which has the following thermodynamic properties along orthogonal directions
\beq \label{dG}
d\mathcal{G}=-sd\mathcal{T}=Pd\mathcal{V}~.
\eeq
From Eq.~\eqref{dG} we conclude that deformations of $\mathcal{G}$ along orthogonal directions cause the material to stretch or compress. After integrating the density \eqref{gibbs_neutral} over the worldvolume one finds the relation
\beq \label{i0}
F[X^{\mu}]=-\int_{\mathcal{W}_{p+1}}\sqrt{-\gamma} \mathcal{G}=-\left(M-\sum_{i}\Omega_{i}J^{i}-TS\right)~.
\eeq
Extremizing \eqref{i0} while keeping the set of potentials $T,\Omega_i$ fixed implies the first law of thermodynamics to be satisfied for the fluid branes:
\beq
dM=\sum_{i}\Omega_i dJ^{i}+TdS~.
\eeq
This formula can interpreted as a prediction \cite{Emparan:2009at}, namely, that a stationary fluid configuration living on a particular dynamical surface must globally satisfy the first law of thermodynamics.

\subsection{Deformations of fluid branes and elastic waves}
The considerations of the previous sections are based on the action \eqref{f0} being only a function of the state of strain given a fixed temperature $T$ and angular velocities $\Omega_i$. Therefore, providing the pressure $P$ as a function of the independent components of the state of strain $\gamma_{ab}$, i.e., $P(\gamma_{ab})$, deformations of the embedding geometry by explicit variation of the embedding map $X^{\mu}(\sigma^{a})$ can be analyzed through variations of the induced metric $\gamma_{ab}$. These take the form \cite{Grignani:2010xm}
\beq
\delta \gamma_{ab}=g_{\mu\nu,\lambda}\partial_a X^{\mu}\partial_{b} X^{\nu}\delta X^{\lambda}+g_{\mu\lambda}(\partial_a X^{\mu}\partial_b\delta X^{\lambda}+\partial_b X^{\mu} \partial_a \delta X^{\lambda})~.
\eeq
Hence, searching for the extrema of the action \eqref{f0} implies that
\beq \label{mot1}
T^{ab}\delta\gamma_{ab}=0~,
\eeq
where the effective stress-energy tensor is obtained in the usual way
\beq\label{offstress}
T^{ab}=\frac{2}{\sqrt{-\gamma}}\frac{\delta I}{\delta \gamma_{ab}}~.
\eeq
For the present case, using \eqref{offstress}, the stress-energy tensor has the general form
\beq \label{ss1}
T^{ab}=2\frac{\partial P}{\partial \gamma_{ab}}+\sigma^{ab}~,
\eeq
and is generally different from $\sigma^{ab}$. Note that \eqref{ss1} is equal to \eqref{pstress} when the relation \eqref{forces} is used. The stress-tensor \eqref{ss1} obtained from the action is only valid along orthogonal directions and can be regarded as an off-shell form of \eqref{pstress} since a priori one can compute \eqref{ss1} without the knowledge of the extrinsic curvature of the embedding. The equations of motion obtained from Eq.~\eqref{mot1} when projected along tangential and orthogonal directions to the worldvolume give rise to the intrinsic and extrinsic equations \eqref{bf_eqs} \cite{Grignani:2010xm} together with the boundary term
\beq
\left[\sqrt{-\gamma}T^{ab}u^{\mu}_{a}\hat{n}_{b}\delta X_{\mu}\right]|_{\partial_{\mathcal{W}_{p+1}}}~,
\eeq
yielding the boundary condition \eqref{bc_thin} and hence vanishing independently of the initial and final configurations by construction. It is also possible to define an off-shell form of the elasticity tensor \eqref{elasticity} using \eqref{ss1}, this is given by
\beq \label{offshellE}
E^{abcd}=2\left(P\gamma^{a(c}\gamma^{d)b}-\left(\frac{\partial P}{\partial \gamma_{ab}}\right)\gamma^{c d}-2\left(\frac{\partial^2 P}{\partial \gamma_{ab}\partial\gamma_{cd}}\right)\right)~.
\eeq
Projecting \eqref{offshellE} using \eqref{sigmatilde} for a particular embedding surface results in \eqref{elasticity}.

\subsubsection*{The speed of elastic waves} 
The dynamical properties of fluid branes as elastic materials can also be seen by applying a small perturbation to the brane geometry. To this aim we follow \cite{Emparan:2009at}. Assuming the material to be initially at rest $u^{a}=(1,0,...)$, we introduce a perturbation in the mapping functions $\delta X^{\mu}$ and initial pressure $P$ such that
\beq
\delta P~,~\delta \rho=\frac{d\rho}{dP}\delta P,~\delta u^{a}=(0,v^{i}),~\delta X^{\mu}=\xi^{\mu}~.
\eeq
Using the form of the stress tensor \eqref{0stress} together with the equation of extrinsic dynamics \eqref{bfe_ext} we find
\beq
\left((\rho-P)\partial^2_t-K\partial_i^2\right)\xi^\mu=0~,
\eeq
and hence conclude that elastic waves propagate at the speed
\beq
c^2_\perp=\frac{K}{\rho-P}~,
\eeq
where we used the definitions of the modulus of rigidity \eqref{mod_rig} and of the solid density \eqref{s_dens}. The above result is what is expected from a relativistic solid which has been subject to hydrostatic compression \cite{PhysRevD.7.1590, Karlovini:2002fc}.


\subsection{Charged fluid branes} \label{c_fluids}
Fluid configurations carrying a $q$-charge are not only characterized by the stress tensor \eqref{st_thin} but also by a totally anti-symmetric current tensor $\hat{J}^{\mu_1...\mu_{q+1}}$ which is confined to the worldvolume surface
\beq \label{j_thin}
\hat{J}^{\mu_1...\mu_{q+1}}(x^{\alpha})=\int_{\mathcal{W}_{p+1}}d^{p+1}\sigma\sqrt{-\gamma}J^{\mu_1...\mu_{q+1}}(\sigma^{a})\frac{\delta^{D}(x^{\mu}-X^{\mu}(\sigma^{a}))}{\sqrt{-g(x^{\alpha})}}
\eeq
and must satisfy current conservation
\beq \label{j_conserv}
\nabla_{\mu_1}\hat{J}^{\mu_1...\mu_{q+1}}=0~,
\eeq
in the absence of external couplings. The set of effective worldvolume equations that result from \eqref{j_conserv} imply that the current is purely tangential 
\beq
J^{\mu_1...\mu_{q+1}}=u^{\mu_1}_{a_1}...u^{\mu_{q+1}}_{a_{q+1}}J^{a_1...a_{q+1}}~,
\eeq
and
\beq \label{wvj_cons}
D_{a_1}J^{a_1...a_{q+1}}=0 ~,~J^{a_1...a_{q+1}}n_{a_1}|_{\partial\mathcal{W}_{p+1}}=0~.
\eeq
We will now apply these equations to the fluid configurations with $q=p$ charge studied in \cite{Emparan:2011hg} and with $q=0$ charge studied in \cite{Caldarelli:2010xz, Emparan:2011hg}.

\subsubsection*{$q=p$ worldvolume charge}
The fluids carrying $q=p$ worldvolume charge studied in \cite{Emparan:2011hg} are characterized by a worldvolume current density $J^{a_1...a_{p+1}}$ of the form
\beq
J=\mathcal{Q}_{p}\hat{V}_{p+1}~,
\eeq
where $\hat{V}_{p+1}$ is the $(p+1)$-volume form of the embedding surface and $\mathcal{Q}_{p}$ is the charge density. The worldvolume conservation equation \eqref{wvj_cons} then implies
\beq \label{constQ}
\partial_{a}\mathcal{Q}_{p+1}=0~.
\eeq
Therefore for $q=p$ charges the charge density $\mathcal{Q}_{p}$ is not allowed to vary along worldvolume directions, hence the total charge of the configuration $Q_p$ equals the charge density $\mathcal{Q}_p$. Given this, the fluid does not carry any additional extra degrees of freedom associated with the charge and as such the Gibbs-Duhem relations presented in \eqref{dp} still hold for the case at hand. In this way, as long as all extrinsic variations are performed while the charge $Q_p$ is kept constant, the results of the previous sections hold. However, it is possible to introduce a chemical potential $\Phi_p$ conjugate to $\mathcal{Q}_{p}$ \cite{Emparan:2011hg}, giving rise to a well defined global quantity 
\beq \label{phiH}
\Phi_\text{H}^{(p)}=-\int_{\mathcal{B}_{p}}dV_{(p)}\Phi_pu^{a}n_{a}~.
\eeq
Using $\Phi_p$ we can define the Gibbs free energy of the fluid as 
\beq \label{gibbs_charged}
\mathcal{G}=\epsilon -\mathcal{T}s-\Phi_p \mathcal{Q}_p=-P-\Phi_p \mathcal{Q}_p~,
\eeq
where in the second equality we have made use of the relations \eqref{dp}. Therefore, for hydrodynamical fluctuations where the charge $Q_p$ is kept constant due to \eqref{constQ} we find
\beq
d\mathcal{G}=-sd\mathcal{T}-\mathcal{Q}_{p}d\Phi_{p}~,
\eeq
and hence for orthogonal variations that keep the potential \eqref{phiH} constant, ie.,
\beq
d\Phi_p=-\Phi_pd\mathcal{V}~,
\eeq
the action can be recast as \eqref{i0} noting that along those directions $d\mathcal{G}=-\mathcal{G}d\mathcal{V}$. From here one can define an electroelastic modulus of rigidity describing the deformation of the charge potential as
\beq \label{rig_ele}
\frac{1}{K_{\text{E}}}=\left(\frac{\partial\mathcal{V}}{\partial \Phi_p}\right)_T =-\frac{1}{\Phi_p}~.
\eeq
As far as the authors are aware, Eq.~\eqref{rig_ele} does not have a classical analog. We suspect that this modulus of electroelastic rigidity is associated with fluctuations of the charge density in transverse directions in the same way as the isothermal permittivity is associated with worldvolume fluctuations \cite{Emparan:2011hg}. We further note that it is possible to define an electroelasticity tensor associated with $\Phi_p$ as in \eqref{elasticity} but its meaning is unclear. A better understanding of this is presently lacking. 

Finally, extremizing \eqref{i0} at fixed $T,~\Omega_{i},~\Phi_\text{H}^{(p)}$ implies the first law of thermodynamics \cite{Emparan:2011hg}
\beq \label{1st_charged}
dM=\sum_{i}\Omega_i dJ^{i}+TdS+\Phi_{\text{H}}^{(p)}dQ_p~,
\eeq
to be satisfied.

\subsubsection*{$q=0$ worldvolume charge}
Stationary fluids carrying a $q=0$ brane charge were analyzed in \cite{Caldarelli:2010xz, Emparan:2011hg} and have quite different thermodynamic properties than the $q=p$ case. These are instead characterized by the worldvolume particle current\footnote{From hereon we omit the index in $\mathcal{Q}_p$ and $\Phi_p$.}
\beq
J^{a}=\mathcal{Q}u^{a}~,
\eeq
and hence must satisfy \eqref{wvj_cons}:
\beq
D_{a}\left(\mathcal{Q}u^{a}\right)=0~.
\eeq
The crucial difference with the $q=p$ case is that now the charge density $\mathcal{Q}$ is allowed to vary along the worldvolume and hence adds extra degrees of freedom to the system. Local thermodynamic equilibrium implies
\beq
d\epsilon=\mathcal{T}ds+\Phi d\mathcal{Q}\,
\eeq
while the thermodynamic Gibbs-Duhem relations \eqref{dp} are now changed to
\beq \label{dp2}
\epsilon+P=\mathcal{T}s+\Phi\mathcal{Q}~~,~~dP=sd\mathcal{T}+\Phi d\mathcal{Q}~.
\eeq
The intrinsic equations of motion \eqref{bfe_int} again lead to conservation of the entropy current as in \eqref{entropy_conser} while the Euler equations \eqref{Euler_force} are modified to \cite{Caldarelli:2010xz}
\beq \label{Euler_force2}
P^{ab}\mathcal{T}s(\dot{u}_b+\partial_b\text{ln}\mathcal{T})-\mathcal{Q}\Phi\left(\hat{K}^{a}-P^{ab}\partial_b \ln\Phi\right)=0~,
\eeq
where we have introduced the mean curvature of the worldlines embedded in $\mathcal{W}_{p+1}$,
\beq
\hat{K}^{a}=u^{b}u^{c}D_{b}\left(u_{c}u^{a}\right)~.
\eeq
Assuming the solution to be stationary, we take the fluid velocities to be aligned with a worldvolume Killing vector field as in \eqref{u}. This ensures that the first term in Eq.~\eqref{Euler_force2} vanishes leading to the same relation between local and global fluid temperatures \eqref{ass_t}. The second term in Eq.~\eqref{Euler_force2} can be dealt with by imposing the mean curvature due to the dissolved $0$-charge on the worldvolume  to balance the gradient of the chemical potential \cite{Caldarelli:2010xz}, yielding
\beq
\hat{K}^{a}=P^{ab}\partial_b\ln \Phi~.
\eeq
On the other hand, the extrinsic equation \eqref{udot} now becomes
\beq
\mathcal{T}s{\perp^{\rho}}_{\nu}\dot{u}^{\mu}+\mathcal{Q}{\perp^{\rho}}_{\nu}\partial^{\nu}\Phi+PK^{\rho}=0~,
\eeq
which after using \eqref{u} and contracting with an arbitrary orthogonal vector $N_\rho$ leads to
\beq
Pd\mathcal{V}+sd\mathcal{T}+\mathcal{Q}d\Phi=0~.
\eeq
This can be integrated to the action \eqref{f0}. In order to make further contact with electroelasticity we define the solid density
\beq \label{newrho}
\rho=\mathcal{T}s+\Phi\mathcal{Q}~,
\eeq
which under deformations along the extrinsic directions satisfies
\beq
d\rho=\mathcal{T}ds-Pd\mathcal{V}+\Phi d\mathcal{Q}~.
\eeq
This is the thermodynamic relation that a solid charged under a particle current should respect. From here, as before, we can obtain useful relations along orthogonal directions:
\beq \label{rel_charged}
\left(\frac{\partial\rho}{\partial s}\right)_{\mathcal{V},\mathcal{Q}}=\mathcal{T}~,~\left(\frac{\partial\rho}{\partial \mathcal{V}}\right)_{s,\mathcal{Q}}=-P~,~\left(\frac{\partial\rho}{\partial\mathcal{Q}}\right)_{s,\mathcal{V}}=\Phi~.
\eeq
Unlike the neutral case, not everything can be derived from a mass function. In fact, if we use the definition \eqref{newrho} into \eqref{rel_charged} we find the identities
\beq
\begin{split}
\left(\frac{\partial\Phi}{\partial \mathcal{T}}\right)_{\mathcal{V},\mathcal{Q}}=\left(\frac{\partial\Phi}{\partial \mathcal{T}}\right)_{s,\mathcal{V}}&=-\frac{s}{\mathcal{Q}} ~,\\
s\left(\frac{\partial\mathcal{T}}{\partial \mathcal{V}}\right)_{s,\mathcal{Q}}+\mathcal{Q}\left(\frac{\partial\Phi}{\partial \mathcal{V}}\right)_{s,\mathcal{Q}}&=-P ~.
\end{split}
\eeq
The Gibbs free energy introduced in \eqref{gibbs_charged} for charged fluids is now, due to \eqref{dp2}, equal to the pressure as in \eqref{gibbs_neutral}. Hence, if one wishes to specify an equation of state of the form
\beq
\epsilon=wP+\Phi\mathcal{Q}~,
\eeq
one finds the action \eqref{solid_action}:
\beq \label{solid_action2}
F[X^{\mu}]=-\frac{1}{w+1}\int_{\mathcal{W}_{p+1}}\sqrt{-\gamma}\left(\rho-\Phi\mathcal{Q}\right)~.
\eeq
When written in terms of the solid density $\rho$, the extra term on the r.h.s. of the action \eqref{solid_action2} acts as an external force. Variation of \eqref{solid_action2} keeping $T,\Omega_i,\Phi_\text{H}$ constant leads to the first law of thermodynamics \eqref{1st_charged}, but now with global charge \cite{Caldarelli:2010xz}:
\beq 
Q=-\int_{\mathcal{B}_{p}}dV_{(p)}\mathcal{Q}u^{a}n_{a}~.
\eeq


\section{Application to stationary black $p$-branes} \label{bf_branes}
In this section we apply the considerations of the previous sections, as an example, to the case of neutral black $p$-branes wrapped on a generic submanifold (blackfolds) studied in \cite{Emparan:2007wm, Emparan:2009at, Emparan:2009vd, Caldarelli:2008pz, Armas:2010hz}. Such branes are characterized by a stress-energy tensor of the type presented in \eqref{st_thin} and hence encompassed within the framework put forth above. We begin by describing the effective blackfold fluid characterizing these branes and then write down the elasticity tensor as well as an action of the thermodynamic type for blackfold objects. We finish by re-deriving the elastic equilibrium condition for higher dimensional black rings and black odd-spheres \cite{Emparan:2009vd}.


\subsection{The effective blackfold fluid}
The blackfold effective fluid characterizing neutral black branes is of the perfect fluid type \eqref{0stress} and characterized by the equation of state \cite{Emparan:2009at}
\beq \label{eq_state}
\epsilon=-(n+1)P~.
\eeq
The local thermodynamic quantities associated with the fluid living on the brane $(\epsilon,~P,~\mathcal{T},~s)$ can all be described in terms of the brane thickness $r_{0}$ in the following way:
\beq \label{black_prop}
\frac{\epsilon}{n+1}=-P=\frac{\Omega_{(n+1)}r_{0}^n}{16\pi G}~~,~~\mathcal{T}=\frac{n}{4\pi r_0}~~,~~s=\frac{\Omega_{(n+1)}r_{0}^{n+1}}{4G}~.
\eeq
Here $n$ is the number of transverse directions given by $n=D-p-3$. By inspection of the above expressions it is easily observed that the Gibbs-Duhem relations \eqref{dp} are satisfied. The thickness $r_0$, as well as the fluid velocities $u^{a}$ are allowed to be functions of $\sigma^{a}$ and hence to vary along worldvolume directions. From \eqref{black_prop} one concludes that the requirement of local thermodynamic equilibrium \eqref{thermo_eq} implies the hierarchy of scales
\beq
r_{0}(\sigma^{a})\ll R(\sigma^a)~,
\eeq
or in other words, that the black branes being wrapped on the submanifold $\mathcal{W}_{p+1}$ are thin compared to the curvature radius of the submanifold.


\subsection{Elasticity tensor of blackfolds and thermodynamic action}
Here we write down the elasticity tensor for neutral blackfolds using Eq.~\eqref{offshellE} and then an action of the thermodynamic type. We begin by noting that from Eq.~\eqref{ass_t} and using the quantities \eqref{black_prop} we obtain a relation between $|\mathbf{k}|$ and $r_0$ 
\beq \label{rel_r0}
r_0=\left(\frac{n}{4\pi T}\right)|\mathbf{k}|=\lambda |\mathbf{k}|~,
\eeq
where we have defined $\lambda=n/4\pi T$. Using this, the pressure $P$ can be expressed in terms of the induced metric $\gamma_{ab}$ as
\beq \label{p_k}
P(\gamma_{ab})=-\frac{\Omega_{(n+1)}}{16\pi G}\lambda^{n}|-\gamma_{ab}\mathbf{k}^{a}\mathbf{k}^{b}|^{\frac{n}{2}}~.
\eeq
Hence, the derivative of the pressure with respect to the induced metric is simply given by
\beq \label{der_p}
\left(\frac{\partial P}{\partial \gamma_{ab}}\right)=-\frac{n}{2}Pu^{a}u^{b}~.
\eeq
Using the expression for the effective stress-energy tensor derived in \eqref{ss1} we arrive at
\beq
T^{ab}=P\left(-nu^{a}u^{b}+\gamma^{ab}\right)~,
\eeq
agreeing with \eqref{0stress} when the equation of state \eqref{eq_state} is introduced. In order to continue further, it is necessary to evaluate the second derivative of $P$ with respect to the state of strain. This yields:
\beq
\left(\frac{\partial^2 P}{\partial \gamma_{ab}\partial \gamma_{cd}}\right)=\frac{n(n-2)}{4}Pu^{a}u^{b}u^{c}u^{d}~.
\eeq
Therefore, using Eq.~\eqref{offshellE} we obtain the off-shell elasticity tensor for neutral blackfolds:
\beq \label{elasticity_black}
E^{abcd}=P\left(2\gamma^{a(c}\gamma^{d)b}+nu^{a}u^{b}\gamma^{cd}-n(n-2)u^{a}u^{b}u^{c}u^{d}\right)~.
\eeq
This is not manifestly invariant under $(a,b)\to (c,d)$ as discussed around Eq.~\eqref{elasticity} but it becomes so after the equations of motion \eqref{ext0} are imposed. We note that \eqref{elasticity_black} has the same structure as the Young modulus tensor measured for black branes in \cite{Armas:2011uf, Camps:2012hw, Armas:2012ac}\footnote{There is a difference between the structure of \eqref{elasticity_black} and the Young modulus measured in \cite{Armas:2011uf, Camps:2012hw, Armas:2012ac}, namely that here there is no ambiguity in the choice of worldvolume surface since we take the surface to be infinitely thin.}. Moreover, it satisfies the properties
\beq
E^{abcd}\gamma_{cd}=P\left(2\gamma^{ab}+n(n+p-1)u^{a}u^{b}\right) \quad , \quad E^{abcd}\gamma_{ab}\gamma_{cd}=(2(p+1)-n(n+p-1))P~.
\eeq

\subsubsection*{Thermodynamic action}
Here we rewrite the action \eqref{f0}, adapted to blackfold objects, in terms of the total entropy \eqref{total_entropy}. Using the Gibbs-Duhem relations \eqref{dp} and also \eqref{eq_state} we obtain the relation
\beq \label{p_rel2}
P=-\frac{1}{n}\mathcal{T}s~,
\eeq
which when introduced into the action \eqref{f0} leads to
\beq \label{therm_act_1}
I[X^{\mu}]=-\frac{T}{n}\int_{\mathcal{B}_{p}}dV_{(p)}su^{a}n_{a}=\frac{T\thinspace S}{n}~.
\eeq
Therefore we see that extremizing the blackfold action at constant temperature and angular velocities is to extremize the black hole entropy.


\subsection{Equilibrium condition for black holes}

In this section we apply the results of the previous sections to re-derive the equilibrium condition of thin black rings and black odd 3-spheres \cite{Emparan:2009vd} in asymptotically flat space parametrized by coordinates $(t,r,\theta,\phi,\psi, x^{i})$. We write the induced metric on the worldvolume as
\beq
\gamma_{ab}d\sigma^{a}d\sigma^{b}=-d\tau^2+R^2\left(d\theta^2+\text{cos}^2\theta d\phi^2+\text{sin}^2\theta d\psi^2\right)~,
\eeq
where $R$ is the radius of the sphere (or the ring). The map onto the ambient space-time is simply
\beq
t=\tau,~r=R,~x^{i}=0~,
\eeq
while the remaining coordinates coincide. The non-vanishing components of the extrinsic curvature are
\beq
{K_{\theta\theta}}^{r}=-R~~{K_{\phi\phi}}^{r}=-R\thinspace \text{cos}^2\theta,~~{K_{\psi\psi}}^{r}=-R\thinspace \text{sin}^2\theta~,
\eeq
while the mean extrinsic curvature vector reads
\beq
K^{r}=-\frac{p}{R}~,~p=1,3~.
\eeq


\subsubsection*{Black rings}
Black rings are described by setting $p=1$ and $\theta=0$ in the formulae above. This leads to the induced metric and extrinsic curvature
\beq
\gamma_{ab}d\sigma^{a}d\sigma^{b}=-d\tau^2+R^2d\phi^2~,~{K_{\phi\phi}}^{r}=-R~,~K^{r}=-\frac{1}{R}~.
\eeq
The worldvolume Killing vector field takes the form
\beq
\mathbf{k}=\partial_\tau+\Omega\partial_\phi~,
\eeq
which we use to write the pressure $P$ as
\beq
P=-\frac{\Omega_{(n+1)}}{16\pi G}\lambda^{n}|-\gamma_{\tau\tau}-\gamma_{\phi\phi}\Omega^2|^{\frac{n}{2}}~.
\eeq
Since the only independent and non-zero component of the strain tensor is $d\gamma_{\phi\phi}$ we can use Eq.\eqref{ext0} to obtain the equilibrium 
\beq \label{forcesring}
-2\left(\frac{\partial P}{\partial \gamma_{\phi\phi}}\right)-\sigma^{\phi\phi}=0,
\eeq
which in turn implies, using the result \eqref{der_p},
\beq \label{ring_0}
\Omega^2R^2=\frac{1}{n+1}~.
\eeq 
This equilibrium condition has been derived previously in \cite{Emparan:2009vd} also using the action \eqref{therm_act_1},  here we have merely performed a different derivation. From Eq.~\eqref{forcesring} we see that there is only one non-vanishing component of strain, hence the only on-shell component of the elasticity tensor is given by \eqref{elasticity_black}:
\beq
E^{\phi\phi\phi\phi}=2\left(\frac{n+1}{n}\right)P\gamma^{\phi\phi}\gamma^{\phi\phi}.
\eeq 


\subsubsection*{Black odd-spheres}
Black odd 3-spheres are described by setting $p=3$. We assume them to be rotating with equal angular velocity $\Omega$ along both directions $(\phi,\psi)$. This allows us to write the worldvolume Killing vector field in the form
\beq
\mathbf{k}=\partial_\tau+\Omega\left(\partial_\phi+\partial_\psi\right)~,
\eeq
and hence the pressure as
\beq
P=-\frac{\Omega_{(n+1)}}{16\pi G}\lambda^{n}|-\gamma_{\tau\tau}-\gamma_{\theta\theta}\Omega^2|^{\frac{n}{2}}~.
\eeq
There are now three non-vanishing components of the strain, $d\gamma_{\theta\theta}, ~d\gamma_{\phi\phi}$ and $d\gamma_{\psi\psi}$. The last two can be expressed in terms of $d\gamma_{\theta\theta}$ in the following way:
\beq
d\gamma_{\phi\phi}=\text{cos}^2\theta d\gamma_{\theta\theta}~,~d\gamma_{\psi\psi}=\text{sin}^2\theta d\gamma_{\theta\theta}~.
\eeq
Using now Eq.~\eqref{ext0} for the only independent component $d\gamma_{\theta\theta}$ we obtain the equilibrium equation
\beq \label{odd_bal_1}
-2\left(\frac{\partial P}{\partial \gamma_{\theta\theta}}\right)-\tilde\sigma^{\theta\theta}=0~,
\eeq
where the effective stress along $d\gamma_{\theta\theta}$ is given by
\beq
\tilde\sigma^{\theta\theta}=P\left(\gamma^{\theta\theta}+\text{cos}^2\theta\gamma^{\phi\phi}+\text{sin}^2\theta\gamma^{\psi\psi}\right)=p\thinspace P \gamma^{\theta\theta}~.
\eeq
Solving Eq.\eqref{odd_bal_1} results in the equilibrium condition
\beq \label{Om_0_odd}
\Omega^{2}R^2=\frac{p}{n+p}~,
\eeq
which has been derived previously in \cite{Emparan:2009vd}. Even though we only considered the cases $p=1,3$, the result \eqref{Om_0_odd} is valid for all $p$ \cite{Emparan:2009vd}. Eq.~\eqref{ext0} could have been solved using any of the other two non-vanishing components of strain $U_{\phi\phi}$,$U_{\psi\psi}$, in fact, it is easy to see that the effective pressure tensor has in this case the generic form $\tilde\sigma^{ab}=\tilde P \gamma^{ab}$ where $\tilde P=p\thinspace P$. Using expression \eqref{elasticity} for the on-shell value of the elasticity tensor we find the non-vanishing components:
\beq
\begin{split}
\tilde E^{\theta\theta\theta\theta}&=2\left(\frac{n+p}{n}\right)\tilde P\gamma^{\theta\theta}\gamma^{\theta\theta}~,~\tilde E^{\phi\phi\phi\phi}=2\left(\frac{n+p}{n}\right)\tilde P\gamma^{\phi\phi}\gamma^{\phi\phi}~,~\tilde E^{\psi\psi\psi\psi}=2\left(\frac{n+p}{n}\right)\tilde P\gamma^{\psi\psi}\gamma^{\psi\psi},~\\
\tilde E^{\theta\theta\phi\phi}&=2\frac{p}{n}\tilde P \gamma^{\theta\theta}\gamma^{\phi\phi}~,~\tilde E^{\theta\theta\psi\psi}=2\frac{p}{n}\tilde P \gamma^{\theta\theta}\gamma^{\psi\psi}~,~\tilde E^{\phi\phi\psi\psi}=2\frac{p}{n}\tilde P \gamma^{\phi\phi}\gamma^{\psi\psi}~.
\end{split}
\eeq


\section{Conclusions} \label{discussion}
Relativistic elasticity theory has been considered in the literature by a number of authors \cite{PhysRevD.1.1013, Salt:1971, Maugin1978, Sokolov:1994mv, Kijowski:1994eq, Karlovini:2002fc, Beig:2002pk}, one of the most influential works being that conducted by Carter and Quintana \cite{Quintana:1972}. The main focus of these works has been on describing elastic solids which are space-filling, contrary to being confined to a dynamical surface. However, the main difference compared to the work presented here is not only related to the existence of an embedding surface but also to the usual assumption regarding the form of the pressure \eqref{pressure_tensor} and elasticity \eqref{elasticity} tensors:
\beq \label{transverse}
\sigma^{ab}u_{b}=0\quad,\quad E^{abcd}u_{d}=0~.
\eeq
This orthogonality condition with respect to the fluid flows seems to have classical roots, namely, that an elastic material can be deformed in space but not in time. Not imposing \eqref{transverse} has been the main point of departure in this note from the usual formulations of relativistic elasticity and has drawn inspiration from the recent results of bending black branes \cite{Armas:2011uf, Camps:2012hw, Armas:2012ac}. 

From the point of view of material science, materials that do not satisfy \eqref{transverse} might belong to a rather unusual class, but from the point of gravitational physics, in which black holes in certain regimes may take the effective description as a fluid brane \cite{Bhattacharyya:2008jc}-\cite{Camps:2012hw}, they are relevant for the analysis of their dynamics. Their study can lead to the formulation of a corner of relativistic elasticity theory that has not been previously explored (e.g. by applying extrinsic perturbations to black brane geometries) in the same way that the fluid/gravity led to a more general formulation of dissipative corrections to fluid and superfluid hydrodynamics \cite{Banerjee:2008th, Erdmenger:2008rm, Son:2009tf, Banerjee:2008th, Herzog:2011ec, Bhattacharya:2011tra}. Here, as an example, we have written down the most general form of the elasticity tensor of a fluid brane that responds only to changes in volume \eqref{elasticity} and computed it for the particular case of neutral blackfolds \eqref{elasticity_black}. The assumptions taken were that the intrinsic properties of the material living on the thin surface were those of a stationary fluid in local thermodynamic equilibrium whose backreaction onto the ambient space-time is negligible. It is unclear, at the present moment, wether taking another kind of intrinsic dynamics would lead to an elastic behavior since it was crucial for our understanding that the extrinsic equations of motion \eqref{ext} could be integrated into an action that only depends on the distances between neighboring points on the brane after fixing the potentials $T,\Omega_{i}, \Phi_{\text{H}}$.

It is not surprising that a stationary fluid will behave elastically along orthogonal directions to the surface on which it lives. Given a perfect fluid with density $\epsilon$ and pressure $P$ living on a two-dimensional sphere embedded in four-dimensional space-time and applying a deformation to the surface such that the radius of the sphere increases by a small amount, then the density and pressure will drop (or increase) due to the increase in volume. This is what we observe here. It will be interesting to apply these ideas and develop the theory when finite thickness corrections to the surface are considered \cite{Armas:2011uf, Armas:2012ac}. Considering such effects allows for probing other types of deformations, such as bending, which require a varying concentration of matter along transverse directions.

\section*{Acknowledgements}
We thank Joan Camps, Jakob Gath, Troels Harmark and Andreas Vigand Pedersen for useful discussions. The work of JA was funded by FCT Portugal grant SFRH/BD/45893/2008. The work of NO is supported in part by the Danish National Research Foundation project ``Black holes and their role in quantum gravity''. We wish to thank the KITP for hospitality during the program ``Bits, branes and black holes'', and to NORDITA during the workshop ``The Holographic Way: String Theory, Gauge Theory and Black Holes".

\appendix

\addcontentsline{toc}{section}{References}
\providecommand{\href}[2]{#2}\begingroup\raggedright\endgroup


\begin{thebibliography}{10}

\bibitem{Damour:1978cg}
T.~Damour, ``{Black Hole Eddy Currents},''
\href{http://dx.doi.org/10.1103/PhysRevD.18.3598}{{\em Phys.Rev.} {\bf D18}
  (1978)  3598--3604}.

\bibitem{Parikh:1997ma}
M.~Parikh and F.~Wilczek, ``{An Action for black hole membranes},''
  \href{http://dx.doi.org/10.1103/PhysRevD.58.064011}{{\em Phys.Rev.} {\bf D58}
  (1998)  064011},
\href{http://arxiv.org/abs/gr-qc/9712077}{{\tt arXiv:gr-qc/9712077 [gr-qc]}}.

\bibitem{Thorne:1986}
K.~Thorne, D.~Macdonald, and R.~Price, ``{Black holes: The Membrane
  paradigm},'' {\em Yale University Press} (1986)  .

\bibitem{Eling:2009sj}
C.~Eling and Y.~Oz, ``{Relativistic CFT Hydrodynamics from the Membrane
  Paradigm},'' \href{http://dx.doi.org/10.1007/JHEP02(2010)069}{{\em JHEP} {\bf
  1002} (2010)  069},
\href{http://arxiv.org/abs/0906.4999}{{\tt arXiv:0906.4999 [hep-th]}}.

\bibitem{Bredberg:2010ky}
I.~Bredberg, C.~Keeler, V.~Lysov, and A.~Strominger, ``{Wilsonian Approach to
  Fluid/Gravity Duality},''
  \href{http://dx.doi.org/10.1007/JHEP03(2011)141}{{\em JHEP} {\bf 1103} (2011)
   141},
\href{http://arxiv.org/abs/1006.1902}{{\tt arXiv:1006.1902 [hep-th]}}.

\bibitem{Bhattacharyya:2008jc}
S.~Bhattacharyya, V.~E. Hubeny, S.~Minwalla, and M.~Rangamani, ``{Nonlinear
  Fluid Dynamics from Gravity},''
  \href{http://dx.doi.org/10.1088/1126-6708/2008/02/045}{{\em JHEP} {\bf 0802}
  (2008)  045},
\href{http://arxiv.org/abs/0712.2456}{{\tt arXiv:0712.2456 [hep-th]}}.

\bibitem{Kovtun:2004de}
P.~Kovtun, D.~Son, and A.~Starinets, ``{Viscosity in strongly interacting
  quantum field theories from black hole physics},''
  \href{http://dx.doi.org/10.1103/PhysRevLett.94.111601}{{\em Phys.Rev.Lett.}
  {\bf 94} (2005)  111601}, \href{http://arxiv.org/abs/hep-th/0405231}{{\tt
  arXiv:hep-th/0405231 [hep-th]}}.
An Essay submitted to 2004 Gravity Research Foundation competition.

\bibitem{Emparan:2007wm}
R.~Emparan, T.~Harmark, V.~Niarchos, N.~A. Obers, and M.~J. Rodriguez, ``{The
  Phase Structure of Higher-Dimensional Black Rings and Black Holes},''
  \href{http://dx.doi.org/10.1088/1126-6708/2007/10/110}{{\em JHEP} {\bf 10}
  (2007)  110},
\href{http://arxiv.org/abs/0708.2181}{{\tt arXiv:0708.2181 [hep-th]}}.

\bibitem{Emparan:2009cs}
R.~Emparan, T.~Harmark, V.~Niarchos, and N.~A. Obers, ``{World-Volume Effective
  Theory for Higher-Dimensional Black Holes},''
  \href{http://dx.doi.org/10.1103/PhysRevLett.102.191301}{{\em Phys. Rev.
  Lett.} {\bf 102} (2009)  191301},
\href{http://arxiv.org/abs/0902.0427}{{\tt arXiv:0902.0427 [hep-th]}}.

\bibitem{Emparan:2009at}
R.~Emparan, T.~Harmark, V.~Niarchos, and N.~A. Obers, ``{Essentials of
  Blackfold Dynamics},'' \href{http://dx.doi.org/10.1007/JHEP03(2010)063}{{\em
  JHEP} {\bf 03} (2010)  063},
\href{http://arxiv.org/abs/0910.1601}{{\tt arXiv:0910.1601 [hep-th]}}.

\bibitem{Camps:2010br}
J.~Camps, R.~Emparan, and N.~Haddad, ``{Black Brane Viscosity and the
  Gregory-Laflamme Instability},''
  \href{http://dx.doi.org/10.1007/JHEP05(2010)042}{{\em JHEP} {\bf 05} (2010)
  042},
\href{http://arxiv.org/abs/1003.3636}{{\tt arXiv:1003.3636 [hep-th]}}.

\bibitem{Camps:2012hw}
J.~Camps and R.~Emparan, ``{Derivation of the blackfold effective theory},''
  \href{http://dx.doi.org/10.1007/JHEP03(2012)038}{{\em JHEP} {\bf 1203} (2012)
   038},
\href{http://arxiv.org/abs/1201.3506}{{\tt arXiv:1201.3506 [hep-th]}}.

\bibitem{Emparan:2011hg}
R.~Emparan, T.~Harmark, V.~Niarchos, and N.~A. Obers, ``{Blackfolds in
  Supergravity and String Theory},''
  \href{http://dx.doi.org/10.1007/JHEP08(2011)154}{{\em JHEP} {\bf 1108} (2011)
   154},
\href{http://arxiv.org/abs/1106.4428}{{\tt arXiv:1106.4428 [hep-th]}}.

\bibitem{Caldarelli:2010xz}
M.~M. Caldarelli, R.~Emparan, and B.~Van~Pol, ``{Higher-dimensional Rotating
  Charged Black Holes},'' \href{http://dx.doi.org/10.1007/JHEP04(2011)013}{{\em
  JHEP} {\bf 1104} (2011)  013},
\href{http://arxiv.org/abs/1012.4517}{{\tt arXiv:1012.4517 [hep-th]}}.

\bibitem{Carter:2000wv}
B.~Carter, ``Essentials of classical brane dynamics,'' {\em Int. J. Theor.
  Phys.} {\bf 40} (2001)  2099--2130,
\href{http://arxiv.org/abs/gr-qc/0012036}{{\tt gr-qc/0012036}}.

\bibitem{Landau:1959te}
L.~Landau and E.~M. Lifshitz, ``Theory of elasticity,''
{\em Course of Theoretical Physics} {\bf Vol. 7} (1959)  134.

\bibitem{Armas:2011uf}
J.~Armas, J.~Camps, T.~Harmark, and N.~A. Obers, ``{The Young Modulus of Black
  Strings and the Fine Structure of Blackfolds},''
  \href{http://dx.doi.org/10.1007/JHEP02(2012)110}{{\em JHEP} {\bf 1202} (2012)
   110},
\href{http://arxiv.org/abs/1110.4835}{{\tt arXiv:1110.4835 [hep-th]}}.

\bibitem{Armas:2012ac}
J.~Armas, J.~Gath, and N.~A. Obers, ``{Black Branes as Piezoelectrics},''
\href{http://arxiv.org/abs/1209.2127}{{\tt arXiv:1209.2127 [hep-th]}}.

\bibitem{Banerjee:2008th}
N.~Banerjee, J.~Bhattacharya, S.~Bhattacharyya, S.~Dutta, R.~Loganayagam, {\em
  et al.}, ``{Hydrodynamics from charged black branes},''
  \href{http://dx.doi.org/10.1007/JHEP01(2011)094}{{\em JHEP} {\bf 1101} (2011)
   094},
\href{http://arxiv.org/abs/0809.2596}{{\tt arXiv:0809.2596 [hep-th]}}.

\bibitem{Erdmenger:2008rm}
J.~Erdmenger, M.~Haack, M.~Kaminski, and A.~Yarom, ``{Fluid dynamics of
  R-charged black holes},''
  \href{http://dx.doi.org/10.1088/1126-6708/2009/01/055}{{\em JHEP} {\bf 0901}
  (2009)  055},
\href{http://arxiv.org/abs/0809.2488}{{\tt arXiv:0809.2488 [hep-th]}}.

\bibitem{Son:2009tf}
D.~T. Son and P.~Surowka, ``{Hydrodynamics with Triangle Anomalies},''
  \href{http://dx.doi.org/10.1103/PhysRevLett.103.191601}{{\em Phys.Rev.Lett.}
  {\bf 103} (2009)  191601},
\href{http://arxiv.org/abs/0906.5044}{{\tt arXiv:0906.5044 [hep-th]}}.

\bibitem{Herzog:2011ec}
C.~P. Herzog, N.~Lisker, P.~Surowka, and A.~Yarom, ``{Transport in holographic
  superfluids},'' \href{http://dx.doi.org/10.1007/JHEP08(2011)052}{{\em JHEP}
  {\bf 1108} (2011)  052},
\href{http://arxiv.org/abs/1101.3330}{{\tt arXiv:1101.3330 [hep-th]}}.

\bibitem{Bhattacharya:2011tra}
J.~Bhattacharya, S.~Bhattacharyya, S.~Minwalla, and A.~Yarom, ``{A Theory of
  first order dissipative superfluid dynamics},''
\href{http://arxiv.org/abs/1105.3733}{{\tt arXiv:1105.3733 [hep-th]}}.

\bibitem{Caldarelli:2008mv}
M.~M. Caldarelli, O.~J.~C. Dias, R.~Emparan, and D.~Klemm, ``{Black Holes as
  Lumps of Fluid},''
  \href{http://dx.doi.org/10.1088/1126-6708/2009/04/024}{{\em JHEP} {\bf 04}
  (2009)  024},
\href{http://arxiv.org/abs/0811.2381}{{\tt arXiv:0811.2381 [hep-th]}}.

\bibitem{Quintana:1972}
B.~Carter and H.~Quintana, ``{Foundations of general relativistic high-pressure
  elasticity theory},'' {\em Proc.Roy.Soc.Lond.} {\bf A331} (1972)  57--83.

\bibitem{Karlovini:2002fc}
M.~Karlovini and L.~Samuelsson, ``{Elastic stars in general relativity. 1.
  Foundations and equilibrium models},''
  \href{http://dx.doi.org/10.1088/0264-9381/20/16/307}{{\em Class.Quant.Grav.}
  {\bf 20} (2003)  3613--3648},
\href{http://arxiv.org/abs/gr-qc/0211026}{{\tt arXiv:gr-qc/0211026 [gr-qc]}}.

\bibitem{Beig:2002pk}
R.~Beig and B.~G. Schmidt, ``{Relativistic elasticity},''
  \href{http://dx.doi.org/10.1088/0264-9381/20/5/308}{{\em Class.Quant.Grav.}
  {\bf 20} (2003)  889--904},
\href{http://arxiv.org/abs/gr-qc/0211054}{{\tt arXiv:gr-qc/0211054 [gr-qc]}}.

\bibitem{PhysRevD.1.1013}
W.~C. Hernandez, \href{http://dx.doi.org/10.1103/PhysRevD.1.1013}{``Elasticity
  theory in general relativity,''{\em Phys. Rev. D} {\bf 1} (Feb, 1970)
  1013--1018}. \url{http://link.aps.org/doi/10.1103/PhysRevD.1.1013}.

\bibitem{Sokolov:1994mv}
S.~Sokolov, ``{Equations of motion of elastic continuum in gravitational
  field},''
{\em IFVE-94-9} (1994)  .

\bibitem{Sokolov:2003}
S.~Sokolov, ``{Elasticity Theory in GR},''
\href{http://arxiv.org/abs/http://web.ihep.su/library/pubs/tconf03/c2-4.htm}{{%
\tt http://web.ihep.su/library/pubs/tconf03/c2-4.htm}}.

\bibitem{Brown:1992kc}
J.~D. Brown, ``{Action functionals for relativistic perfect fluids},''
  \href{http://dx.doi.org/10.1088/0264-9381/10/8/017}{{\em Class.Quant.Grav.}
  {\bf 10} (1993)  1579--1606},
\href{http://arxiv.org/abs/gr-qc/9304026}{{\tt arXiv:gr-qc/9304026 [gr-qc]}}.

\bibitem{Salt:1971}
D.~Salt, ``{Action principles for the elastic solid and the perfect fluid in
  general relativity},''
{\em J. Phys. A.} {\bf 4} (1971)  501.

\bibitem{Grignani:2010xm}
G.~Grignani, T.~Harmark, A.~Marini, N.~A. Obers, and M.~Orselli, ``{Heating up
  the BIon},'' \href{http://dx.doi.org/10.1007/JHEP06(2011)058}{{\em JHEP} {\bf
  1106} (2011)  058},
\href{http://arxiv.org/abs/1012.1494}{{\tt arXiv:1012.1494 [hep-th]}}.

\bibitem{Grignani:2012iw}
G.~Grignani, T.~Harmark, A.~Marini, N.~A. Obers, and M.~Orselli, ``{Thermal
  string probes in AdS and finite temperature Wilson loops},''
  \href{http://dx.doi.org/10.1007/JHEP06(2012)144}{{\em JHEP} {\bf 1206} (2012)
   144},
\href{http://arxiv.org/abs/1201.4862}{{\tt arXiv:1201.4862 [hep-th]}}.

\bibitem{Armas:2012bk}
J.~Armas, T.~Harmark, N.~A. Obers, M.~Orselli, and A.~V. Pedersen, ``{Thermal
  Giant Gravitons},''
\href{http://arxiv.org/abs/1207.2789}{{\tt arXiv:1207.2789 [hep-th]}}.

\bibitem{Camps:2008hb}
J.~Camps, R.~Emparan, P.~Figueras, S.~Giusto, and A.~Saxena, ``{Black Rings in
  Taub-NUT and D0-D6 interactions},''
  \href{http://dx.doi.org/10.1088/1126-6708/2009/02/021}{{\em JHEP} {\bf 0902}
  (2009)  021},
\href{http://arxiv.org/abs/0811.2088}{{\tt arXiv:0811.2088 [hep-th]}}.

\bibitem{PhysRevD.7.1590}
B.~Carter, \href{http://dx.doi.org/10.1103/PhysRevD.7.1590}{``Speed of sound in
  a high-pressure general-relativistic solid,''{\em Phys. Rev. D} {\bf 7} (Mar,
  1973)  1590--1593}. \url{http://link.aps.org/doi/10.1103/PhysRevD.7.1590}.

\bibitem{Emparan:2009vd}
R.~Emparan, T.~Harmark, V.~Niarchos, and N.~A. Obers, ``{New Horizons for Black
  Holes and Branes},'' \href{http://dx.doi.org/10.1007/JHEP04(2010)046}{{\em
  JHEP} {\bf 04} (2010)  046},
\href{http://arxiv.org/abs/0912.2352}{{\tt arXiv:0912.2352 [hep-th]}}.

\bibitem{Caldarelli:2008pz}
M.~M. Caldarelli, R.~Emparan, and M.~J. Rodriguez, ``{Black Rings in
  (Anti)-de{S}itter space},''
  \href{http://dx.doi.org/10.1088/1126-6708/2008/11/011}{{\em JHEP} {\bf 11}
  (2008)  011},
\href{http://arxiv.org/abs/0806.1954}{{\tt arXiv:0806.1954 [hep-th]}}.

\bibitem{Armas:2010hz}
J.~Armas and N.~A. Obers, ``{Blackfolds in (Anti)-de Sitter Backgrounds},''
  \href{http://dx.doi.org/10.1103/PhysRevD.83.084039}{{\em Phys.Rev.} {\bf D83}
  (2011)  084039}, \href{http://arxiv.org/abs/1012.5081}{{\tt arXiv:1012.5081
  [hep-th]}}.

\bibitem{Maugin1978}
G.~A. Maugin, ``Exact relativistic theory of wave propagation in prestressed
  nonlinear elastic solids,'' {\em Annales de l'institut Henri PoincarŽ (A)
  Physique thŽorique} {\bf 28} (1978) no.~2, 155--185.
  \url{http://eudml.org/doc/75976}.

\bibitem{Kijowski:1994eq}
J.~Kijowski and G.~Magli, ``{Relativistic elastomechanics is a gauge type
  theory},''
\href{http://arxiv.org/abs/hep-th/9411212}{{\tt arXiv:hep-th/9411212
  [hep-th]}}.

\end{thebibliography}
\end{document}